\documentclass[10pt,twocolumn,letterpaper]{article}

\usepackage{wacv}
\usepackage{times}
\usepackage{epsfig}
\usepackage{graphicx}
\usepackage{amsmath}
\usepackage{amssymb}

\usepackage{subcaption}
\usepackage{comment}
\usepackage[ruled,vlined]{algorithm2e}
\usepackage{booktabs}       
\newcommand{\tdo}[1]{{}}
\newcommand{\MW}{{\lambda_{\mathrm{metric}}}}

\def\wacvPaperID{572} 

\wacvfinalcopy 

\ifwacvfinal
\def\assignedStartPage{9876} 
\fi

\ifwacvfinal
\usepackage[breaklinks=true,bookmarks=false]{hyperref}
\else
\usepackage[pagebackref=true,breaklinks=true,colorlinks,bookmarks=false]{hyperref}
\fi

\ifwacvfinal
\else
\fi

\begin{document}

\title{Deep Sensory Substitution: Noninvasively Enabling Biological Neural Networks to Receive Input from Artificial Neural Networks}

\author{Andrew Port\\
UC Santa Cruz\\
\and
Chelhwon Kim\\
Leia Inc.\\
\and
Mitesh Patel\\
NVIDIA\\
}

\maketitle

\begin{abstract}
As is expressed in the adage ``a picture is worth a thousand words", when using spoken language to communicate visual information, brevity can be a challenge.  This work describes a novel technique for leveraging machine-learned feature embeddings to sonify visual (and other types of) information into a perceptual audio domain, allowing users to perceive this information using only their aural faculty.  The system uses a pretrained image embedding network to extract visual features and embed them in a compact subset of Euclidean space -- this converts the images into feature vectors whose $L^2$ distances can be used as a meaningful measure of similarity.  A generative adversarial network (GAN) is then used to find a distance preserving map from this metric space of feature vectors into the metric space defined by a target audio dataset equipped with either the Euclidean metric or a mel-frequency cepstrum-based psychoacoustic distance metric. We demonstrate this technique by sonifying images of faces into human speech-like audio.  For both target audio metrics, the GAN successfully found a metric preserving mapping, and in human subject tests, users were able to accurately classify audio sonifications of faces.
\end{abstract}

\section{Introduction}

Many humans have a need or desire for a robust form of sensory feedback. We seek to provide an effective avenue for that feedback through their ears.  The task of mapping information from one sensory domain to another for purposes of human sensory augmentation is known as ``sensory substitution". Such techniques are applicable to those with visual impairments, those in need of feedback from medical or external devices (e.g. prosthetics or robotic limbs), and, more generally, anyone who wishes to augment neurological or biological systems and requires a robust form of feedback.

The question that initially motivated this work was ``Can visual information be communicated to a human through sound as effectively and concisely as it is through light?" In the opinion of the authors of this work, the answer seems to be, not through English but perhaps through another type of audio.  For example, a human with average sight, after a brief (e.g. one second long) glance at a room or face, can describe key features of that room or face.  Communicating this quickly acquired knowledge through English words however would likely take more than a second.  Consider also that some mammals (e.g. bats) can effectively use their auditory systems for navigation in low-light environments. There are examples of humans who have learned to use a form of echolocation as well.\cite{echolocationInHumansOverview}  Additionally, recent advancements in machine learning (i.e. deep learning) have provided methods of effectively embedding high-level visual information in relatively low-dimensional Euclidean space (e.g.  \cite{schroff2015facenet}\cite{netvlad}\cite{disentanglingFeatures} for just a few of the many existing examples).

In this work, we propose an unsupervised neural network architecture that maps visual information into a perceptual audio domain. We imagine this as a system that a user would likely need to spend time to learn to use.  Our proposed model leverages a pre-trained feature embedding (trained in a weakly supervised fashion using triplet loss) that embeds visual information into a metric space where the Euclidean distance between feature vectors gives a notion of visual similarity. We exploit this learned metric space by mapping the feature vectors into audio signals while enforcing that the Euclidean distance between each pair of input feature vectors is equal (up to scaling) to a mel-frequency cepstrum-based psychoacoustic distance of sonifications.  To enforce this geometric preservation, we use a simple term given by Eq.~\ref{eq:metric-loss}.  This loss term is novel as far as the authors are aware, however, due to its simplicity and that it is arguably a natural choice, it seems very likely to have been used in other applications related to  machine learning or computer vision.

Our sonification model's convolutional architecture is a variant of WaveGAN~\cite{wavegan}, a GAN~\cite{goodfellow2014generative} designed to synthesize audio that fits into a given distribution -- that distribution being defined by a dataset of audio files.  

We find that this technique allows the learning of a meaningful sonification between two modalities and demonstrate the technique by mapping images of faces to human speech sounds using a model we refer to as ``Earballs"\footnote{A portmanteau of ``ears" and ``eyeballs"}.  We include results from a human subject test to support the claim that information is preserved in a perceptually meaningful way.


The primary contributions of our work can be summarized as follows.
\begin{itemize}
    \item We propose a novel method of sensory substitution capable of sonifying information extracted by a generic geometric embedding model (e.g. any triplet loss trained feature extraction model) into audio in an information preserving fashion.  To our knowledge, this is the first proposed noninvasive method for enabling biological neural networks to receive arbitrary input from artificial neural networks.
    \item We demonstrate the potential of this technique using the simple case of translating facial feature vectors into (non-lingual) human speech-like audio, verifying the successful preservation of information using quantitative metrics, and evaluating the perceptual accessibility of the information using human subject tests.
\end{itemize}

\section{Related Work}
\subsection{Hearing Visual and Geometric Information}
In 1992, \cite{Meijer1992} proposed a system, vOICe, for sonifying grayscale images (represented in the time-frequency domain) by assigning each row and column of the image to a specific frequency and time, respectively. Essentially, the method imagines that the gray scale image as the spectrogram of a sound, then constructs that sound.  The amplitude of each frequency at each time is determined by the corresponding pixel's intensity.  This idea has been further studied in a variety of works by other researchers (see \url{https://www.seeingwithsound.com/literature.htm} for an extensive list of related works).  Several other approaches to visual-to-audio sensory substitution have been studied which do not make use of machine learning.\cite{EyeMusic,KromoPhone,TheVibe,PSVA,LibreAudioView}  Unfortunately, to our knowledge none of these models are available to the public except for the vOICe.  There are also several existing visual-to-tactile models, \cite{vibroglove,hapbelt,BACHYRITA1969,tacfab} including one noteworthy consumer product, BrainPort\cite{Danilov2005BrainportAA}, an FDA-approved 20x20 electrode-based tactile display that rests on the tongue.

In comparison to these pre-existing sensory substitution systems, our model does not seek to encode or preserve the information of each individual pixel, but instead, higher-level information is extracted by a learned feature embedding.  By focusing on mapping the geometric structure of the feature space into the perceptual audio domain, we can sonify information from a wide range of domains into perceptually meaningful audio.  Our proposed system can also be applied to sonify information from non-visual domains.

We are aware of one other attempt to use machine-learned features for sensory substitution.  In the recent work of \cite{AutoencodingSS}, the author used features from a recurrent autoencoder model known as DRAW \cite{DRAW} to encode hand postures as audio.  Note that DRAW, being a recurrent autoencoder, encodes images as a time series of feature vectors.  During inference, images of hand postures were converted to contours then passed through DRAW to obtain the encoded time series.  The author then evaluated several methodologies for producing sound from these encodings and trained himself to use this audio to recognize 15 hand postures with accuracy above what would be expected by random chance.  This test was performed using a real-time wearable prototype.  \cite{AutoencodingSS} focused on converting each image into a time-series represented some type of contour.  They then investigated multiple ways of converting this time series into audio.  In contrast, we use a pretrained network trained on a large dataset to extract visual information from a color image and encode that information geometrically.  As the information is encoded geometrically (i.e. using triplet loss), we can ensure its preservation as we sonify the features, using the method described in \ref{sec:metric-loss}.


There have also been efforts towards exploring other methods of improving humans ability to perceive their environments through their ears, such as \cite{soniceye}, which investigated the prospect of improving humans' ability to use echolocation through the use of an ultrasonic emitter coupled with a hearing aid to lower the emitter's ultrasonic frequencies into the normal range of human hearing.

As mentioned in the previous section, there have been many attempts at using deep neural networks to perform crossmodal translation tasks, such as the reconstructing frames of video based on video's audio~\cite{crossmodal-duan} or reconstructing images based on English language descriptions of those images~\cite{qi_cross-modal_2018}.  In contrast with these works, our sensory substitution task has the goal of learning a novel map between two domains of distinct modalities, as opposed to learning a pre-existing map or a map that relies on pre-existing interactions between modalities.

\subsection{Geometric Preservation}
Geometric preservation terms have been proposed to improve the efficiency~\cite{onesided}\cite{Fu_2019_CVPR} of GAN models  and also to alleviate mode collapse and the vanishing gradient problem~\cite{distGAN}.  \cite{onesided} used a geometric preservation term similar to Eq.~\ref{eq:metric-loss} for the task of unpaired image to image translation achieving similar results to CycleGAN~\cite{CycleGAN2017} and DiscoGAN~\cite{DiscoGAN} without the need for a secondary inverted generator. 
Our geometric preservation methodology and loss term differ from the previous works we (the authors) are aware of in the choice of normalization method (see Section~\ref{sec:metric-loss}).

\section{Method}
\begin{figure*}[t]
	\centering
	\includegraphics[width=0.8\linewidth]{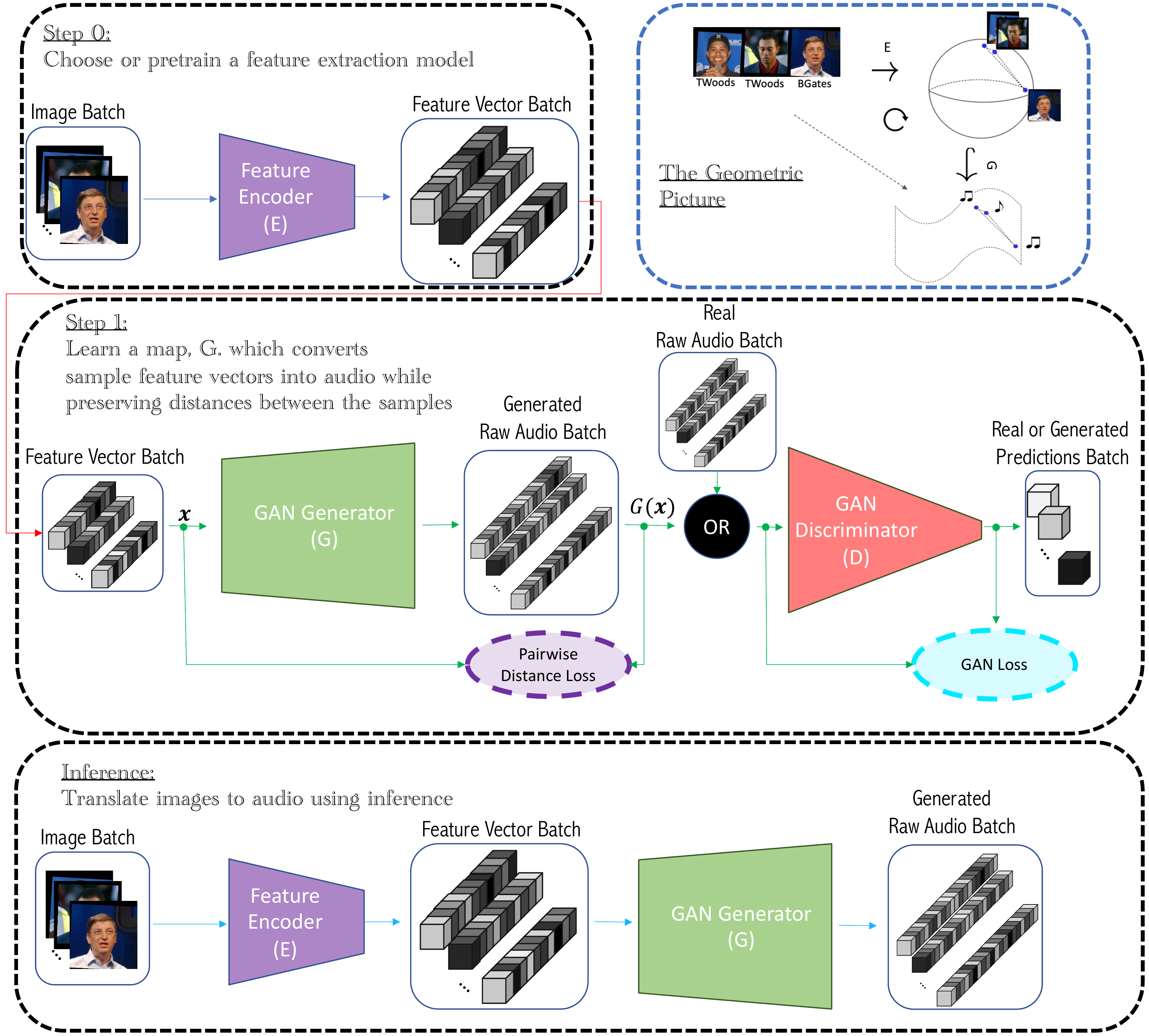}
	\caption{Our proposed image-to-audio sonification model's structure. We use ``OR" here to indicate that the input batch comes from (exclusively) one of the two inputs with equal probability.}
	\label{fig:network}
\end{figure*}


In this work, we propose a deep learning-based framework that translates high-level information extracted from an image or other signal (e.g.  facial identity/expression, location of objects, etc.) into audio. This system can be built on top of any feature embedding model that embeds inputs into a metric space (i.e. any model, $f:X\to Y$, where $d(f(x_1), f(x_2))$ is meaningful for some metric $d$). Our proposed ``Earballs" system (shown in Figure~\ref{fig:network}) starts with a pre-trained feature embedding model capable of extracting desired features from an image (e.g. FaceNet~\cite{schroff2015facenet}). 
We then train a generative network that maps the features into a target perceptual audio domain. This perceptual audio domain can be determined by any sufficiently large and diverse dataset of sounds (e.g. clips of human speech~\cite{panayotov2015librispeech}, musical sounds~\cite{nsynth2017}, etc.).

We use a GAN approach~\cite{goodfellow2014generative} to train the generative network, enforcing that both 1. the output sounds fit into the distribution of sounds specified by the target dataset and 2. that the distances between the output sounds are proportional to the distances between the corresponding inputs samples.

\subsection{WaveGAN}\label{sec:wavegan}
To generate sounds that fit into a target audio dataset, we use a GAN. As is typical with GANs, this is a convolutional neural network composed of two smaller networks trained simultaneously, a generator network and a discriminator network.  The discriminator network is trained as a classifier to differentiate between real samples taken from a target (in our case audio) dataset and samples output by the generator network.  The generator, given a latent input vector, is tasked with generating output that fools the discriminator into predicting it is a real sample from the dataset.

For its generator and discriminator architecture, Earballs uses a variant of WaveGAN\cite{wavegan}, however unlike a typical GAN, during training we alternate between feeding the generator latent input and feature vectors from our source domain.  See Section~\ref{sec:URI} for further explanation of this technique.  We add a novel geometrically motivated loss term, Eq.~\ref{eq:metric-loss}, to enforce that the generator preserves distances between the feature vectors in the input batch.

\subsection{Mel-Frequency Cepstral Coefficients (MFCCs)} \label{sec:mfcc}
The mel scale~\cite{MelScale} is a logarithmic scaling of frequency designed to more closely match human hearing -- the goal being that if humans perceive two frequencies, $f_1$ and $f_2$, as being the same distance apart as two other frequencies, $f_3$ and $f_4$, then $f_1$ and $f_2$ should be the same number of mels apart as $f_3$ and $f_4$, i.e. we should have that ${|m(f_1)-m(f_2)|=|m(f_3)-m(f_4)|}$.  There are multiple implementations of the mel scale, ours follow that used in the TensorFlow signal module: 
\begin{equation}
    m(f) = 1127 \ln\left(1 + \dfrac{f}{700}\right).
\end{equation}

MFCCs are a spectrogram variant that spaces frequency bands on the mel scale.  They are a commonly used audio featurization technique for speech~\cite{MFCC4speech} and musical~\cite{PerceptualDistance} applications.  Our exact implementation uses 80 bins over the frequency range of 80-7600Hz, window length of 1024 samples, fast Fourier transform (FFT) length of 1024, a frame step of 256.

\subsection{Audio Metrics}\label{sec:audio-metric}
For our target domain metric we investigate two metrics, $L^2$ distance, and an MFCC-based metric:
\begin{equation}
    d_Y(y_1, y_2)  = \|\mathrm{MFCC}(y_1) - \mathrm{MFCC}(y_2)\|_2^2
\end{equation}
where $\mathrm{MFCC}$ is as described above in Section~\ref{sec:mfcc}.

\subsection{Geometric Preservation Constraint}\label{sec:metric-loss}
Here we will define our geometric preservation constraint, which encourages our model to learn a mapping that, up to a scaling factor, preserves distances between samples.  Below we will use $G$ to denote our sonification, $(Y, d_Y)$ to denote the target audio domain we are mapping into, equipped with some audio metric, $d_Y$, and $(X, d_X)$ to denote our source domain.  The source domain is any metric space we would like to learn a map into audio for (e.g. a space of feature vectors, points from a 3d model, LiDAR/RGB-D data, etc.).  Note that the information we wish to communicate from $X$ must be contained in the geometry of $(X, d_X)$.  E.g. in the example we’re using as our main demonstration, $X$ is a set of feature vectors output by FaceNet and $d_X(x, x') = \|x - x'\|_2$ is Euclidean distance.  As we used triplet loss to train FaceNet, the entirety of the information we’d like to sonify is contained in these distances.

Let $(X, d_X)$ and $(Y, d_Y)$ be our source and target metric spaces respectively, and let $G: X\to Y$ be the mapping between these spaces given by our generator.  Our metric preservation loss term is then given by
\begin{align}\begin{split}
        &\mathcal{L}_\mathrm{metric}(\mathbf{x}; G) = \label{eq:metric-loss}\\
        &\phantom{\mathcal{L}}{\tfrac{1}{N(N-1)}}\displaystyle\sum_{i < j} \left|\dfrac{d_X(x_i, x_j)}{\mathbb{E}_X[d_X]} - \dfrac{d_Y(G(x_i), G(x_j))}{\mathbb{E}_{G(X)}[d_Y]} \right|
\end{split}\end{align}
where $\mathbf{x} = \{x_1, ..., x_N\} \subset X$ is a batch of $N$ source samples, and $\mathbb{E}_X[d_X]$ and $\mathbb{E}_{G(X)}[d_Y]$ are the mean pairwise distance of samples in the input and in the output batch respectively.

We normalize by the batch means as we want to enforce metric preservation only up to a scaling constant.  This contrasts with previous works (e.g. \cite{onesided}) which standardized over global statistics computed over all of $X$ and $Y$.  This is appropriate in our case for the following reasons -- we will first explain why we normalize instead of standardizing and then why we use batch statistics instead of global statistics.  Conceptually, dividing by the mean can be thought of as scaling both domains to have a mean pairwise distance of one.  Standardization has a similar effect in practice, but comes with greater computational cost and is arguably a less natural choice mathematically.  E.g. consider that for a batch of identical samples, standardization would lead to a divide by zero, whereas dividing by the mean has no such problem.  Standardization is appropriate for scaling a statistical distribution whereas here we want to scale a geometry.  Using global statistics to normalize or standardize these distances would enforce the unnecessary constraint that the image of $G$ has the same diameter as $Y$.  We believe this flexibility can be helpful, which is supported by experimental results shown in rightmost plot in Figure~\ref{fig:metric-preservation-plots} -- our proposed model typically learned a map, $G$, whose image had a smaller diameter than that of $Y$.  

In the example we present below, mapping faces to speech-like sounds, $X$ is a set of 128-dimensional unit vectors output by FaceNet plus some amount of random unit vectors as described in Section~\ref{sec:URI}.  FaceNet is trained using triplet loss, making Euclidean distance the natural choice for $d_X$.  Our $Y$ is given by audio from the TIMIT dataset (preprocessed as described in Section~\ref{sec:TIMIT}). For $d_Y$ we use our MFCC-based audio metric as described in Section~\ref{sec:audio-metric}.

\subsection{Total Loss Functions}
Our generator and discriminator loss functions are, in their totality, given by

\begin{align}
    \begin{split}
        \mathcal{L}_\mathrm{G}&(\mathbf{x}; G, D)\label{eq:g-loss}\\
        &= - D(G(\mathbf{x})) + \MW\mathcal{L}_\mathrm{metric}(\mathbf{x}; G)
    \end{split}\\
    \begin{split}
        \mathcal{L}_\mathrm{D}&(\mathbf{x}, \mathbf{a}; G, D)\label{eq:d-loss}\\
        &= D(G(\mathbf{x})) - D(\mathbf{a}) + \lambda_\mathrm{gp}\mathcal{L}_\mathrm{gp}(\mathbf{x}, \mathbf{a}; G, D)
    \end{split}
\end{align}
where $\mathbf{x}$ is a batch of feature vectors from the source domain, $\mathbf{a}$ is a batch of audio clips from the target audio domain, $G$ and $D$ denote the generator and discriminator respectively, and $\mathcal{L}_\mathrm{gp}$ is a Wasserstein gradient penalty term.  We set $\lambda_\mathrm{gp} = 10$ for all experiments.  We experiment with tuning $\MW$ in Figure~\ref{fig:metric-preservation-plots}.

\section{Results}
In this section, we demonstrate the two ($L^2$ and MFCC) variants of our Earballs model on the task of mapping facial features to human speech-like sounds in a manner useful to the human ear, using WaveGAN without geometric preservation as a baseline.

\subsection{Source Dataset: FaceNet Feature Vectors}\label{sec:LFW}
For our source dataset, we use feature vectors created by the pre-trained OpenFace~\cite{brandon2016openface} implementation of FaceNet~\cite{schroff2015facenet} from images in the Labeled Faces in the Wild (LFW) dataset~\cite{lfw}.  These feature vectors are holistic.  It is worth noting that human facial recognition has been found to be holistic to some significant extent.\cite{VANBELLE20102620}

FaceNet is trained using triplet loss to embed images of the same face close together and images of different faces far apart by Euclidean distance.  This makes $L^2$ distance the natural metric choice for our source domain.

LFW is a collection of 13233 images of 5749 people, 1680 of which have two or more images in the dataset.  The dataset also comes with automatically generated attribute scores for a collection of attributes related to eyewear, hair color, age, etc.

\subsection{Undiscriminated Random Inputs (URI)}\label{sec:URI}
Our proposed model makes use of an augmentation method we will refer to as \emph{undiscriminated random inputs} (URI).  This means that with a probability of $p$ each batch is not taken from the source dataset of feature vectors but instead is randomly uniformly sampled.  The discriminator is not used to evaluate the quality of these samples.  The feature vectors output by our pretrained embedding network being highly clustered, this methodology was implemented in an attempt to help the generator learn the spherical geometry of the input feature vectors.  Potentially, feeding in random feature vectors like this could also help with generalization, enforcing that all possible faces (not just those in the training set) have a unique sound assigned to them.  We analyze the effect of this technique in Figure~\ref{fig:uri-plots} and give some discussion of the results in Section~\ref{sec:audio-fidelity}.
In the case of our FaceNet-based demonstration, our source domain (see Section~\ref{sec:LFW}) is composed of 128-dimensional unit vectors, or equivalently, points on the unit hypersphere in $\mathbb{R}^{128}$.  To uniformly sample points on this sphere, we sample a 128-dimension Gaussian and normalize the samples with respect to Euclidean distance to get unit vectors.  This works with any Gaussian that is symmetric about the origin, as all unit vectors (i.e. all directions) will then be sampled with equal probability density.

\subsection{Target Dataset: TIMIT}\label{sec:TIMIT}
For our target audio dataset, we chose one composed of English human speech sounds, the DARPA TIMIT acoustic-phonetic continuous speech corpus (TIMIT).~\cite{timit}  We chose an English speech dataset thinking that human ears (especially those of native and fluent speakers) might be especially sensitive to these types of sounds and might also find the sounds to be more memorable. The dataset is composed of audio recordings of 630 speakers of eight major dialects of American English, each reading ten ``phonetically rich" sentences.  We randomly shift and then truncate the audio clips to 1.024 seconds (16384 samples at 16kHz).  Note that with our proposed MFCC parameters this gives our target space an effective dimensionality of $61 \times 80 = 4880$ (with respect to geometric preservation).

\subsection{Evaluation Metrics}
We use three computational metrics to evaluate our model as well as two human subject test scores: mean absolute error (MAE), Pearson product-momentum correlation (PC), nearest centroid accuracy (NCA), human subject classification accuracy (HSA), and a human subject memorability score (HSM).  These human subject scores are defined below as well as our NCA metric.  By MAE here we mean exactly our metric loss as defined in Eq.~\ref{eq:metric-loss}.  

\subsubsection{Pearson Product-Momentum Correlation (PC)}\label{sec:pc}
PC is the linear correlation of the pairwise distances of the input samples and the pairwise distances of the output samples, i.e. the dot product of the standardized pairwise distances defined in Eq.~\ref{eq:pc}.  If distances are perfectly preserved, we'd expect a PC of 1.  We'll see that this metric in practice gives us very similar information to that given by MAE -- both metrics vary smoothly together -- however as MAE is also our loss function, we include PC to facilitate the comparison of our model with those using different loss functions. 

\begin{equation}\label{eq:pc}
\begin{aligned}
    &\text{PC}(\mathbf{x}; G) = \\
    &\phantom{\text{PC}}{}\sum_{i < j} 
    \dfrac{
    \left(d_X(x_i, x_j) - \mathbb{E}_X[d_X]\right)
    (d_Y(y_i, y_j) -\mathbb{E}_{G(X)}[d_Y])
    }{\sigma_{dX}\sigma_{dY}}
\end{aligned}
\end{equation}

where here $\mathbf{x} = \{x_1, ..., x_n\}$ is our entire test set of feature vectors, $G:X\to Y$ is our generator, $\mathbb{E}_X[d_X]$ and $\sigma_{dX}$ are the mean and standard deviation of distances between our test set features vectors, and $\mathbb{E}_{G(X)}[d_Y]$ and $\sigma_{dY}$ are the mean and standard deviation of distances between the sonifications of our test set feature vectors.


\subsubsection{Nearest Centroid Accuracy (NCA)}\label{sec:nca}
\begin{figure*}[h]
	\centering
	\begin{subfigure}{1\textwidth}
	\includegraphics[width=\linewidth]{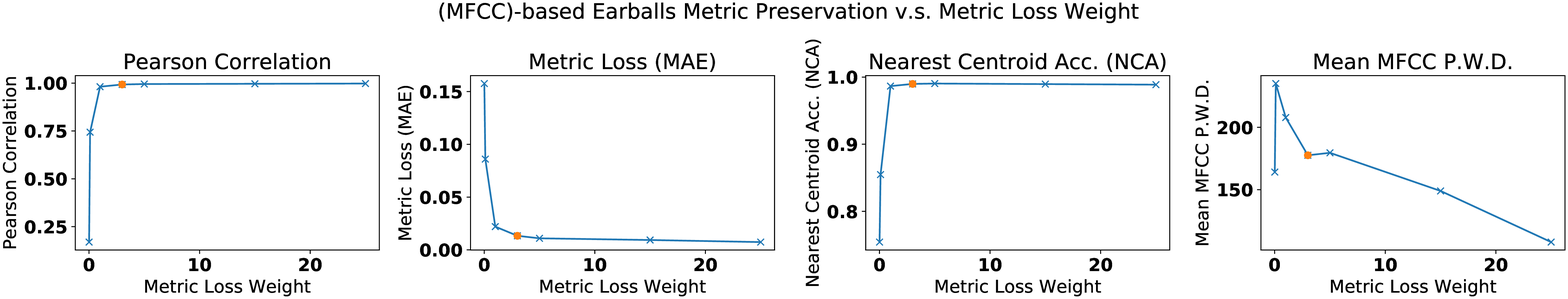}
	\includegraphics[width=\linewidth]{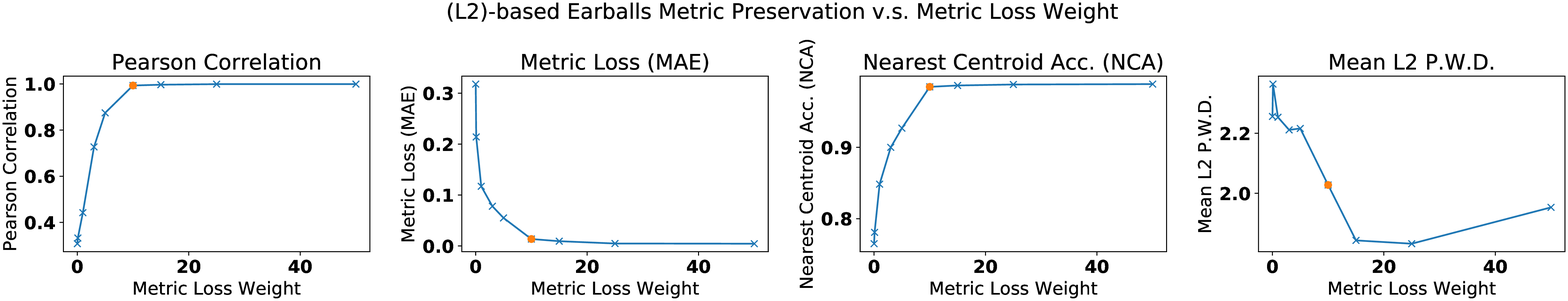}
	\caption{These plots show how our geometric preservation metrics varied with $\lambda_\mathrm{metric}$, the weight of the metric loss term.  In these experiments the source and target domains are, respectively, LFW paired with $L^2$ distance and TIMIT paired with either $L^2$ distance (bottom row) or the MFCC-based audio metric (top row) described in Section~\ref{sec:audio-metric}. 
	The leftmost two plots feature the PC and MAE metrics described above.  The rightmost plot shows the mean pairwise distance between audio sonifications.  Note that the mean pairwise MFCC distance of the TIMIT training set is approximately 200 and the mean pairwise $L^2$ distance is approximately 3.}
	\label{fig:metric-preservation-plots}
	\end{subfigure}
	
	\begin{subfigure}{1\textwidth}
		\includegraphics[width=\linewidth]{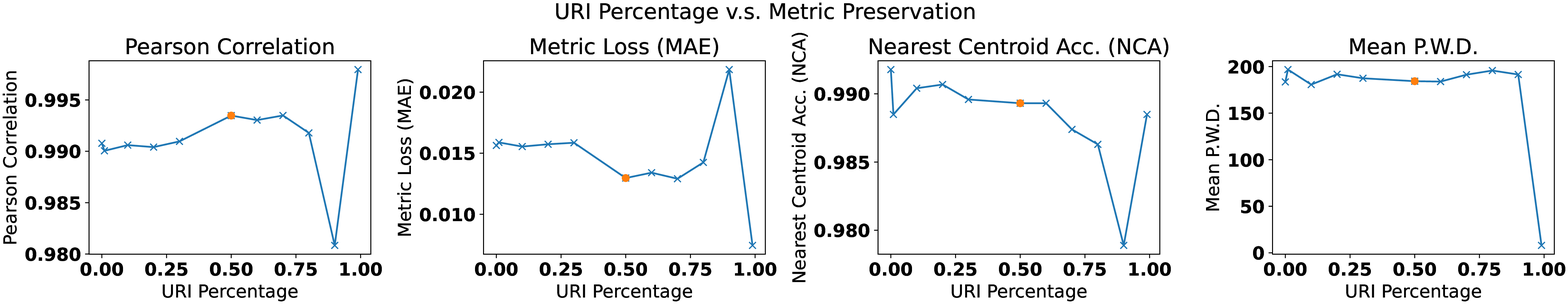}
	\caption{These plots show how our geometric preservation metrics varied with the URI proportion defined in Section~\ref{sec:URI}.  Only the MFCC-based Earballs model was tested in this experiment.  The domains and parameters were otherwise the same as described in Figure~\ref{fig:metric-preservation-plots}.}
	\label{fig:uri-plots}
	\end{subfigure}
\end{figure*}

As our source vectors come from a triplet loss trained model (FaceNet), the source feature vectors in our demonstration can be classified with high accuracy using a nearest centroid classifier.  Using the identity of (the person in the source photo for) each feature vector as its label, all feature vectors (in our test set) with the same label are averaged to find the centroid for each label.  The nearest centroid prediction for any given feature vector is then the label with the closest centroid. The NCA for our test set of FaceNet feature vectors was $\approx 0.948$ before sonification.  As we are focused on preserving the geometry of the source domain, we use the nearest centroid predictions from the source feature vectors as the labels when computing the NCA scores in  Figure~\ref{fig:metric-preservation-plots} -- even if a source vector is nearer to another label's centroid than its own label's, we want to preserve this distance.

\subsection{Human Subject Tests}\label{sec:human-test}
In order to demonstrate that the sonified information can be interpreted by humans, we conducted a human subject test. All participants were students or alumni who responded to a mass email and received a $\$10$ gift card as an incentive.  All tests were administered via one-on-one teleconference by the first author whom we will refer to as the ``administrator" in this section. The administrator was blind to which examples were included in the test and to which of the three models (the MFCC-based Eaballs, $L^2$-based Earballs, or the WaveGAN baseline), was being tested. An institutional review board (IRB) oversight exemption was obtained from the first author's institution's IRB. Subjects were given a 15-minute time limit.  A total of 23 subjects were tested.  Two subjects returned incomplete forms, these data points were removed as the omissions were not noticed until the tests were graded.  The results from the remaining 21 participants are shown in Table~\ref{tab:human-results}.  Before each subject's test began, the test administrator completed a shorter tutorial example of this test (classifying just three samples) themself in front of the subject. The same tutorial example was used for all subjects.

\subsubsection{Human Subject Classification Accuracy (HSA)}\label{sec:hsa}
Participants were provided with three audio clips labeled ``A.wav", ``B.wav", and ``C.wav" and eight additional clips, ``0.wav" through ``7.wav", and asked to classify each as sounding most similar to A, B, or C.  No explanation of ``similar sounding" was given to participants and the tutorial examples completed by the administrator were chosen to be simple to distinguish. These human subject accuracies are reported in Table~\ref{tab:human-results}. A, B, and C were each a sonification of a photo of a different person, each chosen at random from our test set for each participant.  The model used for the test (the MFCC-based Eaballs, $L^2$-based Earballs, or the WaveGAN baseline) was also chosen at random.
In Algorithm~\ref{alg:hs}, we say $L^2$ \emph{separates} categories $\{A, B, C\}$ if no two examples from any of these three categories are farther apart than the closest example from one of the other two categories, where distance here is measure by the $L^2$ distance of the examples' feature vectors.  In cases where the chosen categories were not separable by $L^2$, we generated a new test -- this was appropriate as we were seeking to test the preservation of information and not the accuracy of OpenFace's FaceNet implementation. The probability of a generated test being thrown out in this manner due to inseparability was $\approx68\%$. The tests were generated from examples in our test set in the following manner.

\begin{table*}[t]
  \centering
    \caption{Classification scores ranged from 50\% to 100\%.  The expected accuracy of random guessing is $33.\overline{3}\%$.  The MFCC based model shows a slight advantage with respect to HSA, but significantly lower memorability.  In an earlier round of testing, 7 additional unincentivized participants also evaluated the MFCC-based model receiving an accuracy of 75\%.  HSM was not tested with those participants and those results are not included here.}
  \begin{tabular}{lcccc}
    \toprule
    Method            & Sample Size & HSA & HSA range (out of 8)  & HSM\\
    \midrule
    Earballs (MFCC)   & 9 & 0.667 & 5-8 & 0.778\\
    Earballs ($L^2$)     & 6 & 0.667 & 4-8 & 1.000\\
    WaveGAN (baseline) & 6 & 0.567 & 4-7 & 1.000\\
    \bottomrule
  \end{tabular}

  \label{tab:human-results}
\end{table*}

\begin{table*}
  \centering
    \caption{MAE (Eq.~\ref{eq:metric-loss}), Pearson product-momentum correlation, and mean human subject accuracy for our proposed models and WaveGAN baseline.}
  \begin{tabular}{lcccccc}
    \toprule
                      & \multicolumn{2}{c}{PC}         & \multicolumn{2}{c}{MAE}   & \multicolumn{2}{c}{NCA}                \\
    
    \cmidrule(r){2-3}
    \cmidrule(r){4-5}
    \cmidrule(r){6-7}
    Method            & $L^2$        & MFCC            & $L^2$         & MFCC      & $L^2$      & MFCC\\
    \midrule
    Earballs (MFCC)   & 0.417          & \textbf{0.993}  & 0.226          & \textbf{0.013} & 0.885          & \textbf{0.990}\\
    Earballs ($L^2$)  & \textbf{0.993} & 0.464           & \textbf{0.014} & 0.108          & \textbf{0.985} & \text{0.972}\\
    WaveGAN (baseline) & 0.090          & 0.090           & 0.266534       & 0.567          & 0.856          & 0.664\\
    \bottomrule
  \end{tabular}

  \label{tab:computer-results}
\end{table*}



\begin{algorithm}
\SetAlgoLined
 \While{not done}{
  Sample $0 \leq a, b, c \in \mathbb{Z}$ such that $a + b + c = 8$\;
  Sample categories $\{A, B, C\}$ with more than $1 + \max\{a,b,c\}$ examples each\;
  \If{$L^2$ separates $\{A, B, C\}$}{
  Set done = True\;
  Sample $a+1$, $b+1$, and $c+1$ examples from A, B, and C respectively\;
  }
 }
 \caption{\newline Test~generation~procedure~for~human~subject~tests.}
 \label{alg:hs}
\end{algorithm}


\subsubsection{Human Subject Memorability (HSM)}\label{sec:hsm}
We, the authors, noticed that some of the generated sounds were somewhat memorable -- we would remember them well enough one day to recognize them the next day.  As a rough test for the prevalence of this property among generated samples, in each of the teleconferenced tests, a simple memory test was performed.  After showing the subject the tutorial example, the test administrator would play the user's A, B, or C sound three times.  Directly after this, the test administrator would turn off his camera and microphone and email the participant their test.  The first question on the test response form asks which sound (A, B, or C) was played by the test administrator.  The idea here is to test if the sound was memorable enough that the participant would be able to hold the sound in their head for the approximately 30 seconds that it took to receive the email, unzip the test, and listen to ``A.wav", ``B.wav", and ``C.wav". The results are shown in Table~\ref{tab:human-results}.

\subsection{Training}\label{sec:training}
We train all models for 30k steps using a batch size of 64. This took about 11 hours per run on an NVIDIA 2080 Ti using Adam to optimize both our generator and discriminator with the learning rate, $\beta_1$, and $\beta_2$ set to 0.0001, 0.5, and 0.9 respectively.  The discriminator was updated 5 times for each time the generator was updated.  We experimented with tuning $\MW$, defined in \ref{eq:g-loss}, and the URI proportion (see Figure~\ref{fig:metric-preservation-plots} and Figure~\ref{fig:uri-plots} respectively) finding $\MW = 3$ and a URI proportion of 0.5 to be good choices.  We also experimented extensively with MFCC parameters and other audio metrics, however, we do not provide any related results here as these choices were decided primarily based on personal evaluations of the audio fidelity -- the MFCC parameters and target metric choices greatly affected the quality of the produced audio.  That said, we found the parameters and implementation specifics proposed in this work to consistently lead to good results when using the above-described LFW/FaceNet source domain and TIMIT target domain.

We use LFW's official train-test split which splits the dataset by person, i.e. no person is featured in both the train and test set.  We further split the 4038 person, 9525 sample training set, reserving 415 celebrity identities (1174 samples) for validation.

\subsection{Experimental results}
Tables \ref{tab:computer-results} and \ref{tab:human-results} show our computational and human subject results respectively.  The $L^2$ and MFCC variants of Earballs are competitive.  Results from WaveGAN (without geometric preservation) are included as a baseline.  In Figure~\ref{fig:metric-preservation-plots} we show how three metrics of geometric preservation change as the metric loss weight, $\lambda_\mathrm{metric}$, varies.  The PC, MAE, and NCA improve as $\lambda_\mathrm{metric}$ increases from $\lambda_\mathrm{metric} = 0$ (the baseline case) to  until they plateau around $\lambda_\mathrm{metric}=3$ on our MFCC-based model or $\lambda_\mathrm{metric}=10$ on our $L^2$-based model.  This explains our choices of setting $\lambda_\mathrm{metric} = 3$ and $\lambda_\mathrm{metric} = 10$ for our two Earballs variants.

\subsection{Audio Quality}\label{sec:audio-fidelity}
The generated audio samples might be described as garbed, noisy human speech. There was a trade-off between metric preservation and audio fidelity which can be seen in Figure~\ref{fig:metric-preservation-plots}.   Audio fidelity tended to degrade from the clean sound of a single speaker to a noisy chorus to simply noise in all models as geometric preservation constraints were increased.  As the metric loss weight, $\MW$, increased past about $\MW=5$ the audio quality began to decrease.  By $\MW=100$, the generated audio sounded entirely like noise, not at all like human speech, and sonifications were difficult to discern between by ear.  The parameters recommended above resulted in acceptable audio fidelity without significantly compromising the metric preservation.  This was especially true for the MFCC variant, which by the authors' personal judgment, allowed for increased audio fidelity compared to the $L^2$ variant at the metric preservation plateau point.  We used the URI in both our L2 and MFCC variants because models trained with it tended to have greater audio fidelity. 

\section{Conclusions}
Both variants of Earballs learned to preserve the desired information (as seen by the computational metric results) and much, if not all, of it was preserved in a perceptually usable fashion (as seen by the human subject metric results). We converted faces to sounds as a simple case of sensory substitution.  The results are promising, raising hopes that we might be successful in using the same technique to communicate other high-level visual features through sound.  It may also be possible to communicate lower-dimensional information such as the geometry of a user's surroundings (as viewed through a depth-sensitive camera) or feedback from 2D surfaces (e.g. prosthetics). We picture a practical sensory substitution system to be a tool that a user, at least in the near future, would need to spend significant time to learn to use.  That said, the reward for learning to use such a system could be great, and the system itself could be quite inexpensive, possibly for some applications, even implementable using the sensors and computing power available on a modern smartphone.

{\small
\bibliographystyle{ieee_fullname}
\bibliography{refs}
}

\end{document}